\documentclass{article}
\usepackage{spconf}
\usepackage{graphicx}
\usepackage{bm}
\usepackage{booktabs}
\usepackage{subfig}
\usepackage[font=footnotesize]{caption}
\usepackage{hyperref}
\usepackage{algorithm2e}
\usepackage{nicefrac}
\usepackage{siunitx}
\usepackage{textcomp}

\usepackage{widebar}
\usepackage{localmath}

\graphicspath{{figures/}}

\begin{document}
\ninept

\title{SURROGATE SOURCE MODEL LEARNING FOR DETERMINED SOURCE SEPARATION}

\name{Robin Scheibler and Masahito Togami%
}
\address{LINE Corporation, Tokyo, Japan}

%

\maketitle

\begin{abstract}
  We propose to learn surrogate functions of universal speech priors for determined blind speech separation.
  Deep speech priors are highly desirable due to their high modelling power, but are not compatible with state-of-the-art independent vector analysis based on majorization-minimization (AuxIVA), since deriving the required surrogate function is not easy, nor always possible.
  Instead, we do away with exact majorization and directly approximate the surrogate.
  Taking advantage of iterative source steering (ISS) updates, we back propagate the permutation invariant separation loss through multiple iterations of AuxIVA.
  ISS lends itself well to this task due to its lower complexity and lack of matrix inversion.
  Experiments show large improvements in terms of scale invariant signal-to-distortion (SDR) ratio and word error rate compared to baseline methods.
  Training is done on two speakers mixtures and we experiment with two losses, SDR and coherence.
  We find that the learnt approximate surrogate generalizes well on mixtures of three and four speakers without any modification.
  We also demonstrate generalization to a different variation of the AuxIVA update equations.
  The SDR loss leads to fastest convergence in iterations, while coherence leads to the lowest word error rate (WER).
  We obtain as much as \SI{36}{\percent} reduction in WER.
\end{abstract}


\section{Introduction}

Speech recordings are, as a matter of fact, routinely corrupted by copious amounts of background noise and competing sources.
Source separation offers an attractive way to isolate each of these sounds before further processing, e.g., for automatic speech recognition (ASR)~\cite{Makino:2018iq}.
As a consequence, the topic has recently attracted a considerable amount of attention.
%
A popular and powerful method for speech separation is time-frequency domain masking~\cite{Hershey:2015ve,Erdogan:2015bt}.
Permutation invariant training (PIT) is the default method to train such separation networks~\cite{Kolbaek:2017ct}.
These methods can be extended to the multichannel case by using the masks to estimate spatial statistics of the sources and do beamforming~\cite{Higuchi:2016gq,Heymann:2016gq}.
Another approach uses the spatial cues as input to the network~\cite{Wang:2018hz}.
Deep unfolding proposes to unroll conventional iterative algorithms, such as expectation-minimization, and to learn parameters with backpropagation~\cite{Wisdom:2016gp}.
Recently, several works propose to learn the network parameters by directly optimizing the output of the separation~\cite{Togami:2019mc}.
They extend the method to unsupervised learning~\cite{Togami:2020wc} and resource constrained environments~\cite{Togami:2021c6}.
We note that these latter methods make use of powerful, yet computation-hungry, spatial filtering techniques, limiting the number of iterations of the algorithms through which backpropagation can be safely done.

In this work, we study the training and performance of DNNs as source models for independent vector analysis (IVA).
IVA is a maximum likelihood (ML) approach that considers joint distributions of the sources in the time-frequency domain~\cite{Kim:2006bd,Hiroe:2006ib}.
AuxIVA is an efficient algorithm for IVA relying on majorization-minimization for the optimization~\cite{Ono:2011tn}.
This approach introduces a surrogate function of the true contrast function that can be efficiently minimized.
The quality of the source model in IVA is crucial to the performance of the separation.
However, the choice of the source model is limited to those that have a surrogate function, i.e. super-Gaussian distributions~\cite{Ono:2011tn}, or introduce extra parameters that need to be estimated, e.g., non-negative low-rank models~\cite{Kitamura:2016vj}.
Nevertheless, several recent approaches are based on deep models.
For example, a network can be trained to denoise spectrograms and used in place of a source model~\cite{Nugraha:df,Makishima:2019fl}.
Other approaches train a generative source model on clean data, and then plug it into the likelihood function~\cite{Kameoka:2019be,Li:2020vt}.
All of these approaches use source models tailored to specific musical instruments~\cite{Nugraha:df,Makishima:2019fl} or speakers~\cite{Kameoka:2019be,Li:2020vt}.
In addition, current approaches based on generative models require expensive backpropagation through, or Monte-Carlo sampling of, the source model at inference time~\cite{Kameoka:2019be,Li:2020vt}.

Our approach is straightforward.
We propose to approximate the surrogate function of the true source model with a DNN\@.
Surmising that correct separation indicates a correct source model, we learn the parameters of the DNN by backpropagating through up to 20 iterations of AuxIVA.
This is enabled by iterative source steering (ISS), simple and matrix inverse free update equations for AuxIVA~\cite{Scheibler:2020ig}.
We evaluate two losses applied with PIT to the output of the separation.
End-to-end learning with the scale-invariant signal-to-distortion-ratio (SI-SDR) directly in the time-domain, and coherence, as proposed in a similar work based on the natural gradient algorithm~\cite{Li:2020vt}.
Our approach has the advantage to learn a universal model of single speech sources.
Because the IVA model is independent of the number of speakers, we can easily scale up the separation task to more speakers or even switch the optimization algorithm used for IVA\@.
We demonstrate that even though the learning was done on two speakers mixture, our source model performs equally well on mixtures of three and four speakers.
It also performs well when we replace ISS updates by another type of update equations.
We conjecture this to be possible because the training forces the source model to only learn characteristics of the sources it separates.
In all cases we show large improvements over conventional AuxIVA source models~\cite{Ono:2011tn,Kitamura:2016vj} and some mask-based methods~\cite{Erdogan:2015bt,Heymann:2016gq}.
In addition to SI-SDR, we evaluate the word and character error rates (WER and CER, respectively) for a pre-trained ASR system.
We find reductions of WER of over \SI{20}{\percent} for two speakers, and \SI{30}{\percent} for three and four speakers compared to the next best baseline.

\section{Background}
\seclabel{background}

We consider mixtures of sources captured by $M$ microphones in the short-time Fourier transform (STFT) domain,
\begin{equation}
  \vx_{fn} = \mA_f \vs_{fn} + \vb_{fn}, \quad \forall f,n,
\end{equation}
where $\vx_{fn}\in\C^M$ is the vector containing the $M$ microphone signals at frequency bin $f$ and frame $n$.
The mixing matrix $\mA_f\in\C^{M\times M}$ models the transfer function between the sources, with signals contained in vector $\vs_{fn}\in\C^M$, and the microphones.
Note that we consider the determined case where the number of sources and microphones is the same, i.e.~$M$.
The frequency bin and frame indices $f$ and $n$ run from one to $F$ and $N$, respectively.
In the rest of this manuscript, we denote vectors and matrices by bold lower and upper case letters, respectively.
Furthermore, $\mA^\top$ and $\mA^\H$ denote the transpose and conjugate transpose, respectively of matrix $\mA$.

\subsection{Auxiliary Function based Independent Vector Analysis}

IVA separates the sources with frequency-wise \textit{demixing} matrices,
  $\vy_{fn} = \mW_f \vx_{fn}$, $\forall f,n,$
where $\mW_f \in \C^{M\times M}$.
Since both $\vy_{fn}$ and $\mW_f$ are unknown, this is an ill-posed problem.
A popular way to solve it is by maximum likelihood with a generative model for the sources.
Let $\mY_k \in \C^{F\times N}$ be the complex STFT representation of the $k$th source, i.e. 
\begin{equation}
  (\mY_k)_{fn} = (\vy_{fn})_k = y_{kfn} = \vw_{kf}^\H \vx_{fn},
\end{equation}
where $\vw_{kf}^\H$ is the $k$th demixing filter, e.g., the $k$th row of $\mW_f$.
Then, provided an appropriate prior for the source distribution $p(\mY_k)$, we may recover $\mW_f$ by minimizing the negative log-likelihood,
\begin{equation}
  \ell(\calW) =  \sum_{k=1}^M G(\mY_k) + 2 N \sum_f \log |\det \mW_f|,
\end{equation}
where $\calW$ represents the set of all demixing matrices, and $G(\mY) = - \log p(\mY)$ is the so-called \textit{contrast function}.
Direct minimization of $\ell(\calW)$ is difficult and the MM approach has been shown to be an effective way to do the job, resulting in AuxIVA~\cite{Ono:2011tn} and related methods~\cite{Scheibler:2020ig,Scheibler:2020lqpqm}.
It introduces a surrogate function $G^+(\mY, \wh{\mY})$,
\begin{equation}
  G(\mY) \leq G^{+}(\mY, \wh{\mY}) = \sum_{fn} u_{fn}(\hat{\mY}) |y_{fn}|^2 + c(\wh{\mY}).
\end{equation}
with properties $G(\mY) \leq G^{+}(\mY, \wh{\mY})$, and $G(\wh{\mY}) = G^{+}(\wh{\mY}, \wh{\mY})$, for any $\mY, \wh{\mY}$.
If those properties are satisfied, then choosing the next iterate as the solution to the following problem decreases the cost function $\ell(\calW)$,
\begin{multline}
  \underset{\calW}{\min}\ \sum_{k,f,n} u_{fn}(\mY_k^{(t-1)})|\vw_{kf}^\H \vx_{fn}|^2
  - 2N \sum_f \log|\det \mW_f |,
  \nonumber
\end{multline}
where $t$ is an iteration index.
In the case of $M=2$, this equation can be solved exactly by generalized eigenvalue decomposition (AuxIVA-IP2)~\cite{Ono:2012wa}.
Of particular interest is ISS~\cite{Scheibler:2020ig}, whereas the following substitution is done in the above minimization,
\begin{align}
  \mW_f \gets \mW_f^{(t-1)} - \vv_{kf} (\vw_k^{(t-1)})^H,
\end{align}
and the minimization is done with respect to $\vv_{kf}\in \C^M$ instead, for $k=1,\ldots,M$, in order.
The closed-form solution is
\begin{align}
  (\vv_{kf})_m & = \begin{cases}
    \frac{\sum_n r_{mfn}\, y_{mfn}^{(t-1)}\,\left(y_{kfn}^{(t-1)}\right)^*}{\sum_n r_{mfn} \, |y_{kfn}^{(t-1)}|^2} & \text{if $m\neq k$} \\
    1 - \left(\frac{1}{N} \sum_n r_{mfn} |y_{kfn}^{(t-1)}|^2\right)^{-\nicefrac{1}{2}} & \text{if $m=k$}
  \end{cases}
  \elabel{iss_update}
\end{align}
where $r_{mfn} = u_{fn}\left(\mY_m^{(t-1)}\right)$.
These updates have a simple structure well-suited to implementation in a DNN framework.

\subsection{Source Priors for AuxIVA}
\seclabel{source_priors}

AuxIVA is quite flexible, but requires the source priors to admit a surrogate function.
Here are a few examples of conventional models.

\textit{Circularly symmetric priors} consider distributions over the STFT frame vectors, i.e., $G(\mY) = \sum_n S\big(\big[\sum\nolimits_f |y_{fn}|^2\big]^{\nicefrac{1}{2}}\big)$, where $S$ is a function over the positive real numbers.
If $S$ is a super-Gaussian function, then a surrogate function is readily available~\cite{Ono:2011tn}.
This class includes so-called Laplace with $S(r) = r$, or time-varying Gaussian with $S(r) = -2\log r$.

\textit{Local Gaussian with low-rank}, such as used in ILRMA~\cite{Kitamura:2016vj}, $G(\mY) = \sum_{fn} \frac{|y_{fn}|^2}{\lambda_{fn}}$, where $\lambda_{fn} = \sum_{b=1}^B t_{fb} q_{bn}$, with $t_{fb},q_{bn}\geq 0$, and $b$ small.
An extra step is required to estimate these parameters.

\textit{Deep priors} have been proposed due to their superior modelling power.
Variational autoencoders and other generative models have been used to model different speakers~\cite{Kameoka:2019be,Li:2020mlsp}.
These methods have the drawback that most of them require that samples be drawn from the distribution, which is computationally expensive.

\section{Learning the Surrogate Source Model}

We propose to exploit the modelling power of DNNs to find a good approximation of the surrogate function.
Effectivelly, we replace $u_{fn}(\wh{\mY})$ in \eref{iss_update} by a DNN.
Since the \textit{true} $G(\mY)$ is unknown, and complicated, we will use the separation operation as a proxy to learn the surrogate instead.
Provided that our approximation of $G(\mY,\wh{\mY})$ is good enough, then AuxIVA should provide the best estimate of the separated sources in the maximum likelihood sense.
Thus, we expect it to be close to the \textit{true} surrogate $G(\mY, \wh{\mY})$ after training.

The learning of the model parameters is done by gradient descent, as follows.
The sample is transformed to the STFT domain.
Then, it is run through a fixed number of AuxIVA-ISS iteration with the current model parameters.
After that, the performance of the separation is evaluated using one of the two losses described in \sref{losses}.
For a time-domain loss, the separated sources need to be rescaled as described in \sref{scaling}.

\subsection{Network Architecture of the Approximate Surrogate}
\seclabel{network_arch}

We design our network architecture so as to be able to model medium range dependencies within the time-frequency representation of speech.
To this end, we propose to use a network based on GLU blocks~\cite{Dauphin:2017ft}.
They have been shown to focus on important information in the signal.
The GLU block and network architecture are shown in \ffref{network_arch}.
All convolutions operate along the time axis.
All filters are of length three.
First, we compute the log-magnitude of the complex input.
Then, the first block reduces the number of frequencies to 128 bands.
This is followed by another two GLU blocks with equal input and output sizes, with a dropout layer in the middle.
Finally, a transposed convolution layer brings back the number of frequencies to match that of the input.

\begin{figure}
  \centering
  \includegraphics[width=0.7\linewidth]{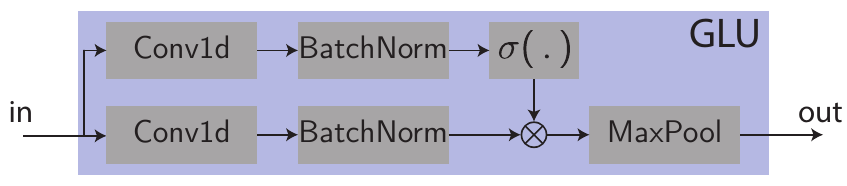}
  
  \vspace{0.25cm}
  \includegraphics[width=0.7\linewidth]{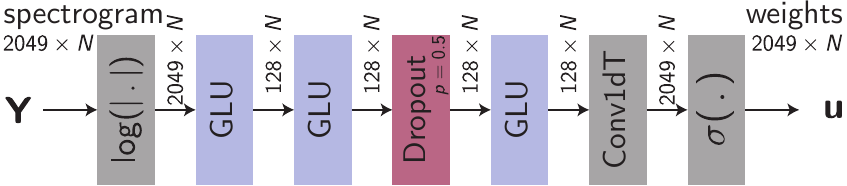}
  \caption{Top: The gated linear unit (GLU) block structure~\cite{Dauphin:2017ft}. Bottom: Details of the deep neural network architecture used.}
  \flabel{network_arch}
\end{figure}

\subsection{Scaling}
\seclabel{scaling}

One of our goals is to apply loss functions directly on the waveform in the time-domain.
However, separation by IVA has a scale ambiguity for the separated signals.
This can be resolved by matching the scale to that of the input signal via the minimal distortion principle~\cite{Matsuoka:2002jy}.
Practically, we find one scalar weight $z_{kf}\in\C$ per source and frequency, minimizing $\E[| x_{mfn} - z_{kf} y_{kfn} |^2]$.
The closed-form solution to this problem is differentiable and can be plugged at the output of the network, before inverse STFT.

\subsection{Loss Functions}
\seclabel{losses}

We explore the use of two cost functions.
Being the most popular metric to judge source separation systems, the SI-SDR~\cite{LeRoux:2018tq} is a prime candidate for direct optimization, as in prior work in the time-domain~\cite{Luo:cd}.
Let $\hat{\vy}$ and $\vs$ be vectors containing the estimate and clean reference signal, respectively, of the $k$th source in the time-domain.
Then, the SI-SDR is
\newcommand{\alphaweight}{\frac{\hat{\vy}^\top \vs}{\vs^\top\vs}}
\begin{align}
  L_{\mathsf{SDR}}(\hat{\vy}, \vs) = 10\log_{10}\left( \frac{\left\|\alpha(\hat{\vy}, \vs)  \vs \right\|^2 }{\left\| \alpha(\hat{\vy}, \vs) \vs - \hat{\vy}\right\|^2} \right),\quad
  \alpha(\hat{\vy}, \vs) = \alphaweight.
  \nonumber
\end{align}
The second metric is the total coherence as proposed in~\cite{Li:2020vt}.
This loss is defined in the STFT domain, with $\wh{\mY}$ and $\mS$ being the STFT of $\hat{\vy}$ and $\vs$, respectively.
Then, the coherence is defined as
\begin{align}
  L_{\mathsf{Coh}}(\hat{\mY}, \mS) = \frac{1}{F} \sum_f \frac{|\E[(\wh{\mY})_{fn} (\mS)_{fn}^*]|}{\sqrt{\E[|(\wh{\mY})_{fn}|^2] \E[|(\mS)_{fn}|^2]}}.
  \elabel{loss_coherence}
\end{align}
Because the order of the sources of the output is ambiguous, we use PIT~\cite{Kolbaek:2017ct},
\begin{align}
  L_{\mathsf{PIT}} = \max_{\pi} \frac{1}{M} \sum_{m} L(\mY_{\pi(m)}, \mS_m),
\end{align}
where $L$ may be either of $L_{\mathsf{SDR}}$ or $L_{\mathsf{Coh}}$, and $\pi$ is taken over all permutations of integers from 1 to $M$.

\section{Experiments}

Our experiments have several goals.
First, we evaluate the performance gap between AuxIVA with a trained model and several baselines.
We do the evaluation both in terms of SI-SDR and SI-SIR~\cite{LeRoux:2018tq}, and WER and CER of an ASR system trained using the \textit{wsj0} recipe from the ESPnet framework~\cite{Watanabe:2018gy}.
Because the \textit{same} model $G(\mY)$ is used for all the sources to separate, and the updates of the demixing matrices do not require any trainable parameters, our proposed method can accommodate any number of speakers.
We want to verify that the source model performs equally well on mixtures of more than two speakers, which were not seen during training.
Thus, in addition to two speakers, we also run the test on mixtures of three and four speakers.
Finally, we want to evaluate how well the model trained with ISS performs when used in a different algorithm.
We want to verify that the training does not specialize the source model to be used only with ISS.
To this end, we repeat the experiment using the model trained for ISS in the IP2 algorithm~\cite{Ono:2012wa} that uses the generalized eigenvalue decomposition (GEVD)\@.
We do this for mixtures of two speakers only.
All the results were obtained from the test data not seen during training, nor used for validation.
All the audio signals are sampled at \SI{16}{\kilo\hertz}.
The STFT frame size is 4096 with half-overlap, and uses a Hamming window.
The number of iterations of AuxIVA is fixed to \numlist{20;50;80} for \numlist{2;3;4} channel mixtures, respectively.

\begin{figure}
  \centering
  \includegraphics[]{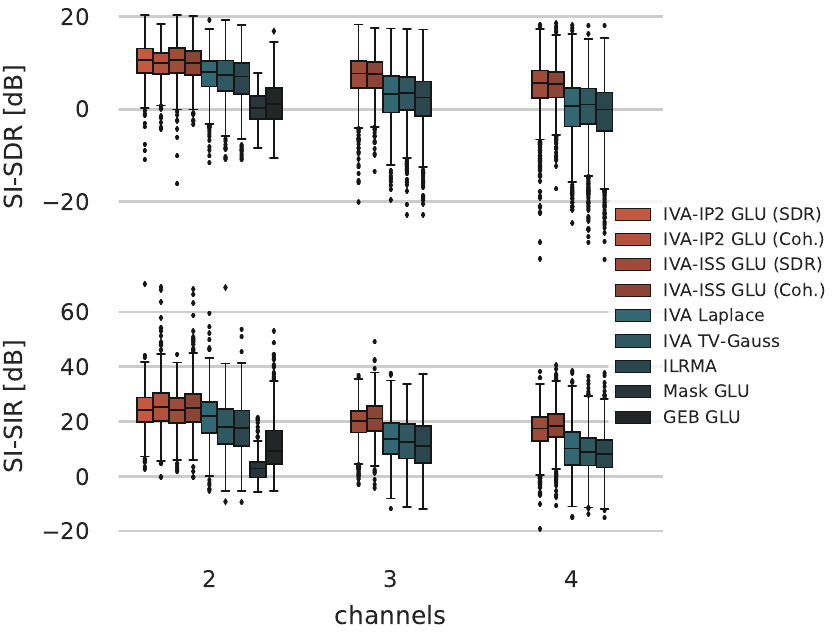}
  \caption{Boxplots of the SI-SDR and SI-SIR after separation. The trained IVA prior is denoted GLU. Mask means single-channel separation, and GEB is the mask-based beamforming.}
  \flabel{sdr_boxplots}
\end{figure}

The baseline methods are as follows.
\textit{1) Conventional AuxIVA models.} These are the source models conventionally used with AuxIVA as described in \sref{source_priors}.
Namely, Laplace and time-varying Gauss circularly symmetric priors~\cite{Ono:2011tn}, and ILRMA~\cite{Kitamura:2016vj}.
For two sources, we use the fast IP2 rules~\cite{Ono:2012wa}.
For two and three sources, we use ISS.
\textit{2) Single channel mask-based separation}~\cite{Erdogan:2015bt}.
We learn a network to produce phase sensitive masks that allow to separate two sources from a single mixture.
The network architecture used is very similar to that presented in~\sref{network_arch} with two differences.
First, instead of the log-magnitude spectrogram, we concatenate the real and imaginary parts of the complex spectrogram before feeding them to the network.
Second, we use two  transposed convolutional layers in parallel to produce the two masks.
\textit{3) Mask-based GEVD beamforming (GEB)}~\cite{Heymann:2016gq}.
Using the masks produced as explained above and all the input channels, we estimate the covariance matrix of each source, as well as the corresponding noise covariance matrix.
This lets us compute the maximum SINR beamformer for each of the sources, which are then used to perform linear separation.

All the algorithms are implemented in Pytorch.
We train all the models on two speakers mixtures from the dataset.
All the mixtures are trimmed to a suitable length for training.
For the optimization, we use Adam.
Due to the iterative nature of AuxIVA, there is a risk of exploding gradient that we mitigate using autoclip~\cite{seetharaman2020autoclip}.
Autoclip sets the gradient clipping parameter to the $p$th percentile of the gradient norms seen so far.
We train all models for at least 30 epochs and pick the one with the best validation SI-SDR.

\begin{figure*}
  \centering
  \includegraphics[]{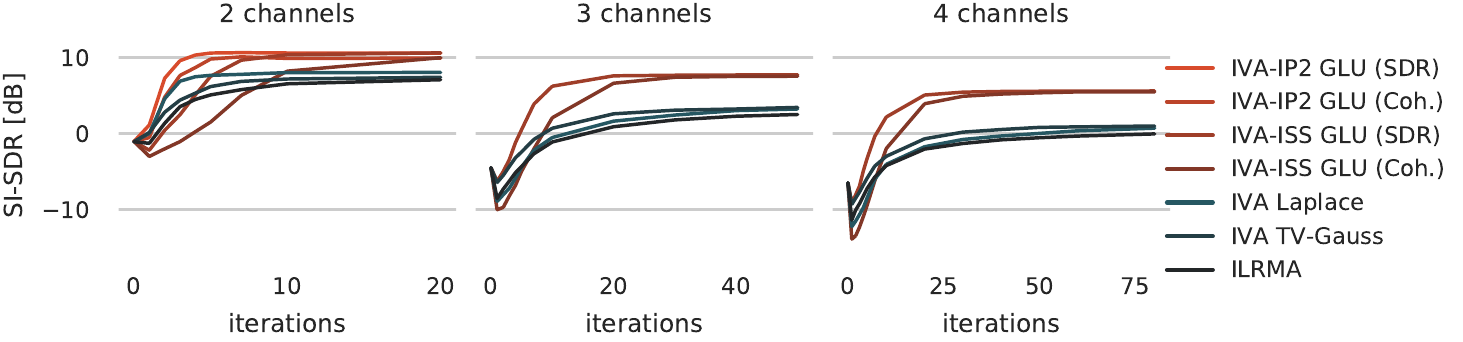}
  \caption{Convergence of the SI-SDR as a function of the number of iterations for the different algorithms.}
  \flabel{sdr_convergence}
\end{figure*}

\subsection{Dataset}
\seclabel{dataset}

We use a dataset of reverberant noisy speech mixtures simulated using \texttt{pyroomacoustics}~\cite{Scheibler:2018di}.
Rooms are created at random following approximately the procedure for the spatialized \textit{wsj0-2mix}~\cite{Wang:2018hz}.
Room walls are between \SI{5}{\metre} and \SI{10}{\metre}, with reverberation time chosen uniformly at random in the \SIlist{200}{\milli\second} to \SI{600}{\milli\second} interval.
Arrangement of microphones and sources was selected at random.
The average power of the first source is normalized to one.
The relative SNR of other sources to the first is chosen at random from \SI{-5}{\decibel} to \SI{5}{\decibel}.
Speech samples were extracted from the WSJ1 corpus~\cite{wsj1}.
Noise from the CHIME3 dataset~\cite{Barker:2015km} was added to the samples to attain an SNR between \SI{10}{\decibel} and \SI{30}{\decibel}.
We randomly shuffle the channels of the CHIME3 multichannel noise recording before adding them to the signal.
Note that the microphone placements in our dataset do not match those of the CHIME3 array.
This procedure can be repeated for varying numbers of sources and microphones.
In the experiments, we use datasets with equal number sources and microphones, namely, two, three, and four of each.
Finally, each of the datasets obtained is split into training, validation, and test sets containing \num{37416}, \num{503}, and \num{333} mixtures, respectively.
As usual, training data is used to train the model, validation to monitor performance and adjust hyperparameters, and test was only run once to obtain the results presented here.

\subsection{Results}

\ffref{sdr_boxplots} shows box-plots of the final SI-SDR and SI-SIR values on the test set.
The median values are given in~\tref{results_sdr_asr}.
IVA generally performs well, but the proposed GLU-based source models outperform all other methods by at least \SIlist{2.6;4.5;4.9}{\decibel} for \numlist{2;3;4} sources, respectively.
Thus, source models trained on two speakers mixtures generalize well to mixtures of more speakers.
The final performance using ISS and IP2 updates is also the same.
In terms of SI-SDR, models using it as a loss perform slightly better, unsurprisingly.
Interestingly, the coherence loss leads to higher SI-SIR than the SI-SDR loss.
Here we find a better performance of IVA based methods than mask based methods.

The convergence speed of AuxIVA is of considerable practical interests.
\ffref{sdr_convergence} shows the evolution of the SI-SDR with the number of iterations.
The learnt source models outperform traditional models both in terms of convergence speed and final performance.
Although the number of iterations is always set to 20 during training on two sources, the convergence is in fact faster than that.
IP2 is known to have faster convergence than ISS, thanks to more effective updates, and it is reassuring to see that this also the case here.
We also find that the coherence loss leads to slower convergence.

Last, but not least, as shown in \tref{results_sdr_asr}, the trained source models lead to a dramatic improvement of WER and CER compared to the best next performing method, i.e. AuxIVA with Laplace model.
For two speaker mixtures, while the \SI{2.6}{\decibel} improvement seemed modest, it leads to over \SI{20}{\percent} and \SI{13}{\percent} reduction in WER and CER, respectively.
The gains are even larger for three and four speaker mixtures, with over \SI{32}{\percent} reduction in WER in both cases.
For this task, the model trained using the coherence loss performs markedly better by up to \SI{3.6}{\percent} reduction in WER compared to the SI-SDR loss for three speakers mixtures.
We conjecture this to be due to the SI-SDR favoring large amplitude elements, at the expense of semantically meaningful, but low amplitude, segments.

\begin{table}
  \centering
  \caption{Median SI-SDR and SI-SIR in decibels, and WER and CER of ASR of the separated signals from the test set. GEB is the mask-based beamforming algorithm~\cite{Heymann:2016gq}, PSM is phase sensitive mask training~\cite{Erdogan:2015bt}, and Coh. is the coherence loss~\eref{loss_coherence}.}
  \begin{tabular}{@{}rlllrrrr@{}}
    \toprule
    Ch. & Algo. & Model     & Loss    & SDR   & SIR  & WER    & CER      \\
    \midrule
    2   & GEB   & GLU       &  PSM    &  1.2  &  9.2 & 95.0\% & 60.5\%   \\
        & IVA   & Laplace   &  --     &  8.1  & 21.9 & 54.5\% & 31.6\%   \\
        & IVA   & GLU       &  SDR    & 10.7  & 24.1 & 33.5\% & 18.0\%   \\
        & IVA   & GLU       &  Coh.   & 10.0  & 24.9 & 33.0\% & 17.8\%   \\
    \midrule
    3   & IVA   & Laplace   &  --     &  3.2  & 13.6 & 80.0\% & 50.3\%   \\
        & IVA   & GLU       &  SDR    &  7.7  & 20.1 & 47.1\% & 27.3\%   \\
        & IVA   & GLU       &  Coh.   &  7.6  & 21.1 & 43.5\% & 25.2\%   \\
    \midrule
    4   & IVA   & Laplace   &  --     &  0.7  & 10.2 & 91.2\% & 58.6\%   \\
        & IVA   & GLU       &  SDR    &  5.6  & 17.4 & 58.3\% & 35.0\%   \\
        & IVA   & GLU       &  Coh.   &  5.5  & 18.4 & 55.3\% & 32.5\%   \\
    \bottomrule
  \end{tabular}
  \tlabel{results_sdr_asr}
\end{table}

\section{Conclusion}
\seclabel{conclusion}

We have proposed to learn surrogate source models for source separation based on the AuxIVA algorithm.
Unlike most previous methods for determined separation with DNN based source models, we learnt the source model directly by backpropagating through AuxIVA the permutation invariant loss on a separation task.
We evaluate two losses for training, SI-SDR applied on the time-domain output signal, and a coherence measure in the time-frequency domain.
In experiments, we show that the learnt source models outperform conventional methods of determined source separation.
Despite the training being done on two speaker mixtures, the performance translated to mixtures of three and four speakers.
We found that the source model generalizes well to a different demixing matrix update algorithm, which means the model learnt may be used for different tasks.
Finally, we showed dramatic improvements of the word error rate of speech recognition using learnt source models.
For this task, the coherence loss produced markedly better results than the SI-SDR, which should be a guiding principle for the evaluation of separation algorithms in the future.


\bibliographystyle{IEEEtran}
\bibliography{IEEEabrv,refs}

\end{document}